\documentclass[aps,amssymb,showpacs,amsmath,superscriptaddress,showkeys,prl, twocolumn]{revtex4-1}
\usepackage[utf8x]{inputenc}
\usepackage{graphicx}
\usepackage{epsf}
\usepackage{natbib}
\usepackage{units}
\usepackage{ulem}
\usepackage{lmodern}  

\newcommand{\reffig}[1]{Fig.~\ref{#1}}
\newcommand{\refeq}[1]{Eq.~\ref{#1}}

\newcommand{\PP}[1]{probe particle }

\usepackage{color}

\begin{document}

\title{Submolecular resolution by variation of IETS amplitude and its relation to AFM/STM signal}

\author{Bruno de la Torre}
\email[corresponding author: ]{bdelatorre@fzu.cz}
\affiliation{Institute of Physics, Academy of Sciences of the Czech Republic, v.v.i.,  Cukrovarnick\' a 10, 162 00 Prague, Czech Republic}
\affiliation{ Regional Centre of Advanced Technologies and Materials, Palack\'{y} University, Olomouc, Czech Republic.}

\author{Martin \v{S}vec}
\affiliation{Institute of Physics, Academy of Sciences of the Czech Republic, v.v.i.,  Cukrovarnick\' a 10, 162 00 Prague, Czech Republic}
\affiliation{ Regional Centre of Advanced Technologies and Materials, Palack\'{y} University, Olomouc, Czech Republic.}

\author{Giuseppe Foti}
\affiliation{Institute of Physics, Academy of Sciences of the Czech Republic, v.v.i.,  Cukrovarnick\' a 10, 162 00 Prague, Czech Republic}

\author{Ond\v{r}ej Krej\v{c}\'{i}}
\affiliation{Institute of Physics, Academy of Sciences of the Czech Republic, v.v.i.,  Cukrovarnick\' a 10, 162 00 Prague, Czech Republic}
\affiliation{Charles University, Faculty of Mathematics and Physics, Department of  Surface and Plasma Science, V Hole$\check s$ovi$\check c$k\'ach 2, 180 00, Prague, Czech Republic}

\author{Prokop Hapala}
\affiliation{Institute of Physics, Academy of Sciences of the Czech Republic, v.v.i.,  Cukrovarnick\' a 10, 162 00 Prague, Czech Republic}

\author{Aran Garcia-Lekue}
\affiliation{Donostia International Physics Center (DIPC), Paseo Manuel Lardizabal 4, E-20018 San Sebastian, Spain}
\affiliation{Ikerbasque, Basque Foundation for Science, 48013 Bilbao, Spain}

\author{Thomas Frederiksen}
\affiliation{Donostia International Physics Center (DIPC), Paseo Manuel Lardizabal 4, E-20018 San Sebastian, Spain}
\affiliation{Ikerbasque, Basque Foundation for Science, 48013 Bilbao, Spain}

\author{Radek Zbo\v{r}il}
\affiliation{ Regional Centre of Advanced Technologies and Materials, Palack\'{y} University, Olomouc, Czech Republic.}

\author{Andres Arnau}
\affiliation{Donostia International Physics Center (DIPC), Paseo Manuel Lardizabal 4, E-20018 San Sebastian, Spain}

\author{H\'{e}ctor V\'{a}zquez}
\affiliation{Institute of Physics, Academy of Sciences of the Czech Republic, v.v.i.,  Cukrovarnick\' a 10, 162 00 Prague, Czech Republic}

\author{Pavel Jel\'{i}nek}
\email{jelinekp@fzu.cz}
\affiliation{Institute of Physics, Academy of Sciences of the Czech Republic, v.v.i.,  Cukrovarnick\' a 10, 162 00 Prague, Czech Republic}
\affiliation{ Regional Centre of Advanced Technologies and Materials, Palack\'{y} University, Olomouc, Czech Republic.}
\affiliation{Donostia International Physics Center (DIPC), Paseo Manuel Lardizabal 4, E-20018 San Sebastian, Spain}


\begin{abstract}
Here we show scanning tunnelling microscopy (STM), non-contact atomic force microscopy (AFM) and inelastic electron tunnelling spectroscopy (IETS) measurements on organic molecule with a CO-terminated tip at 5K. The high-resolution contrast observed simultaneously in all channels unambiguously demonstrates the common imaging mechanism in STM/AFM/IETS, related to the lateral bending of the CO-functionalized tip. The IETS spectroscopy reveals that the submolecular contrast at 5K consists of both renormalization of vibrational frequency and variation of the amplitude of IETS signal. This finding is also corroborated by first principles simulations. We extend accordingly the probe-particle AFM/STM/IETS model to include these two main ingredients necessary to reproduce the high-resolution IETS contrast. We also employ the first principles simulations to get more insight into different response of frustrated translation and rotational modes of CO-tip during imaging.

\end{abstract}


\maketitle

The development of high-resolution scanning tunnelling microscopy (STM) \cite{Temirov_NJP2008}, atomic force microscopy (AFM) \cite{Gross_Science2009} and inelastic electron tunnelling spectroscopy (IETS) \cite{Chiang_Science2014} imaging with functionalized tips allowed to reach unprecedented spatial resolution of organic molecules on surfaces. Using these techniques, the chemical structure of molecules can be now routinely determined directly from experimental images \cite{Gross_NatureChem2010,Schuler_JACS2015}, as well as the information about bond order \cite{Gross_Science2012}, intermediates and products of on-surface chemical reactions \cite{deOteyza_Science2013} or charge distribution within molecules \cite{Hapala_NatCom2016}.

The origin of the high-resolution AFM/STM imaging is now
well understood within the framework of the so-called
probe-particle (PP) model
\cite{Hapala_PRB2014,Hapala_PRL2014,Krejci_PRB2017}. In general,
an~atom or molecule (the probe particle) placed at the tip apex is
sensitive to spatial variations of the potential energy landscape
of the molecule resulting from the interplay between Pauli,
electrostatic and van der Waals interactions
\cite{Moll_NJP2012,Hamalainen_PRL2014,Hapala_PRB2014,Hapala_PRL2014}.
At close tip-sample distances, the probe particle relaxes
according to the potential energy surface, which gives rise to a
sharp submolecular contrast. Nevertheless, neither direct
experimental evidence nor a unified description of the imaging
mechanism for all three scanning modes have been presented.

On one hand, non-contact AFM is most often used in the
high-resolution imaging mode, which provides the highest spatial
resolution with relatively straightforward interpretation,
compared to the other two modes. This implies that the instrument
operates in the frequency modulation (FM) mode
\cite{JAP1991_Albrecht}, which is not a trivial task from both the
instrumental and data acquisition point of view.
From this perspective, the STM mode seems to be a~more feasible
choice. On the other hand, the interpretation of high-resolution
STM images is not at all straightforward because of the
convolution of the geometric and electronic effects
\cite{Krejci_PRB2017}. The IETS mode \cite{Stipe_Science98} thus
represents a~promising alternative \cite{Chiang_Science2014}, but
the high-resolution contrast was so far only demonstrated at
sub-Kelvin temperatures. This temperature requirement poses severe
limitations for its wider application.

In this Letter, we present simultaneous AFM/STM/IETS measurements
of iron(II) phtalocyanine (FePc) on Au(111) acquired with CO-terminated tip at 5~K
\cite{SOM}. These measurements (i) demonstrate that
high-resolution IETS imaging is also feasible with standard
LHe bath cryostats, and (ii) experimentally confirm the common
imaging mechanism for all three imaging modes. Atomistic
simulations using non-equilibrium Green's functions (NEGF) \cite{Frederiksen_PRB2007} and an
extension of the PP model
\cite{SOM,Chiang_Science2014,Hapala_PRL2014} provide a
characterization of the contrast in the inelastic signal and a
unified description of AFM/STM/IETS.

\begin{figure}
\centering
\includegraphics[width=8.5cm]{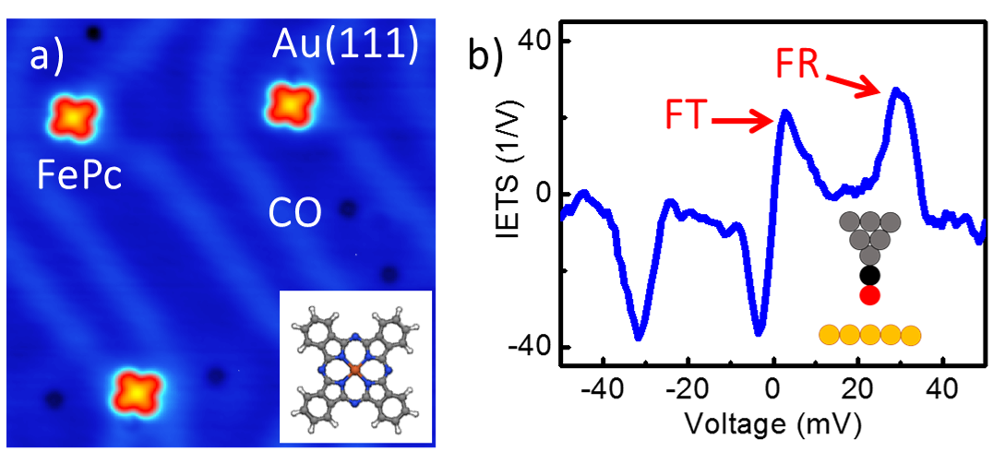}
\caption{\textbf{Constant current STM images of FePc on Au(111)
and IETS signal of CO-functionalized tip.} (a) 16~nm~$\times$~16~nm
constant-current STM image of the Au(111) surface and the
co-adsorbed CO and FePc molecules, acquired at -200~mV and 10~pA
set-point. (b) STM-IETS of the CO adsorbed on the tip apex taken
at the bare Au surface, showing the vibrational modes. The
frustrated rotational (FR) and frustrated translational (FT) modes
of the CO are resolved. Bias voltage modulation was 3~mV at 963~Hz
frequency. Stabilization set point V = 50.0~mV and I = 3.0~nA. }
\label{fig-01}
\end{figure}

\reffig{fig-01}~a) shows a constant-current STM image of
co-adsorbed FePc and CO molecules on Au(111). Prior to the
high-resolution imaging, a CO molecule was picked up to the tip.
The presence of the CO on the tip is confirmed by the
characteristic low-energy IETS spectrum over the bare substrate,
consisting of the frustrated translational (FT) and frustrated
rotational (FR) modes located at $\approx$~3~meV and
$\approx$~30~meV, respectively, as  shown in \reffig{fig-01}~b).
These values are similar to the previous IETS
spectra of a CO molecule adsorbed on Ag(110) or Cu(111)
\cite{Chiang_Science2014,Xu_PRL2016,Okabayashi_PRB2016}.  In
repeated attempts, we noticed that the amplitude and the shape of
the FT peak in the IETS spectrum of the CO tip are very sensitive
to the configuration of the metal tip-apex before its
functionalization \cite{Xu_PRL2016}. For this reason, the metal
apex preparation and CO picking process were repeated until an
intense and regular IETS spectrum was obtained.

\begin{figure}
\centering
\includegraphics[width=8.5cm]{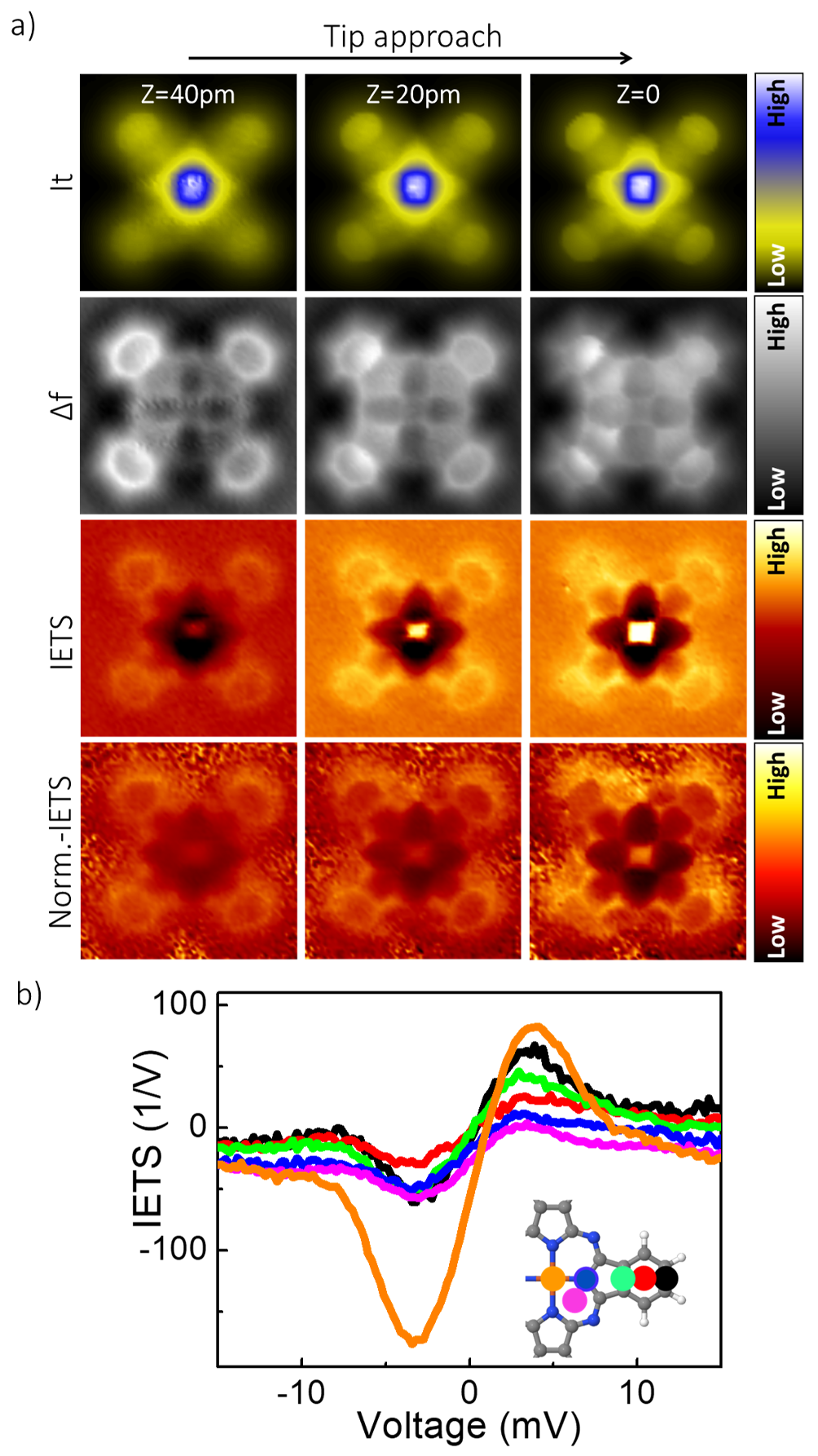}
\caption{\textbf{Simultaneous AFM/STM/IETS images of FePc/Au(111)
with CO-functionalized tip and spatial variation of the FT mode.}
(a) Set of simultaneous (1.6~nm~$\times$~1.6~nm) current, $\Delta$f, IETS
and normalized-IETS constant-height images acquired at three
different tip-sample distances with a CO decorated tip. The images
were acquired with bias voltage of 3.0~mV . Both $\Delta$f and IETS
images show the sharp edges related with the geometric structure
of the FePc molecule. (b) Spatial dependence of the
frustrated translational mode for the CO-functionalized tip above
the FePc molecule.}
 \label{fig-02}
\end{figure}

In \reffig{fig-02}~a) we present the maps of signals obtained
while scanning with the CO-terminated tip in three constant
heights above the molecule. These are the tunnelling current
I$_t$, frequency shift $\Delta$f, $d^2I/dV^2$ and its normalised
value - $(d^2I/dV^2)/(dI/dV)$, denoted in the following as STM,
AFM, IETS and Norm.-IETS respectively. The bias voltage was set to 3~mV in order to
optimise the sensitivity of the IETS signal to the peak
corresponding to the FT vibrational mode. The STM contrast is
dominated by a~strong signal in the central part of the molecule,
suppressing significantly the submolecular resolution of the
molecule. We attribute the strong signal in the central part to
$d$-orbitals of Fe atom just below the Fermi level.

On the other hand, both AFM and IETS images exhibit sharp edges that reveal the backbone of the FePc molecule. At the closest distances the AFM contrast inverts and the contrast in the IETS map is enhanced. The $\Delta$f(Z) spectroscopy shown in Fig.~S1 \cite{SOM} proves that the images were recorded at tip-sample separations where the repulsive interaction plays a dominant role. These observations are fully consistent with the assertion that the characteristic sharp edges in the AFM images are the direct consequence of the lateral bending of the CO due to the repulsive interaction \cite{Hapala_PRB2014}. Since the apparent positions of the sharp edges found in the IETS map correspond almost exactly to the AFM and STM images taken at the same moment, we can infer that the effect of CO bending indeed plays an indispensable role in the IETS imaging as it was predicted theoretically \cite{Hapala_PRL2014}.

Next, we focus on the spatial variation of the FT signal
over the FePc molecule. In their seminal work, Chiang \textit{et al}.
\cite{Chiang_Science2014} performed spatially-resolved IETS
measurements at 600~mK above Co-phtalocyanine molecules. The high
spectral resolution allowed to resolve a subtle variation of the
FT vibrational energy (frequency) and clearly showed a
submolecular contrast. Later a theoretical explanation of the
imaging mechanism was suggested, based on the frequency change of
the mode \cite{Hapala_PRL2014}. However, the IETS-spectra acquired
for the FT mode at various locations of the FePc molecule at 5~K,
shown in the \reffig{fig-02}~b), reveal that the skeleton of FePc can be
clearly resolved purely from variation of the amplitude of the
IETS signal, even though variation of frequency cannot be directly
observed due to broadening of the peak. 

Remarkably, the highest amplitude is detected above the center of the FePc molecule, similarly to the STM dI/dV images. The IETS signal is expected to scale with the magnitude of the differential conductance dI/dV measured at the energy of the FT mode. Indeed, in the normalised IETS images (d$^2$I/dV$^2$)/(dI/dV) the signal in the center is suppressed and the molecular structure is still clearly resolved although the normalization impairs the signal-to-noise ratio in the outer regions of the molecule.

\begin{figure}
\centering
\includegraphics[width=8.5cm]{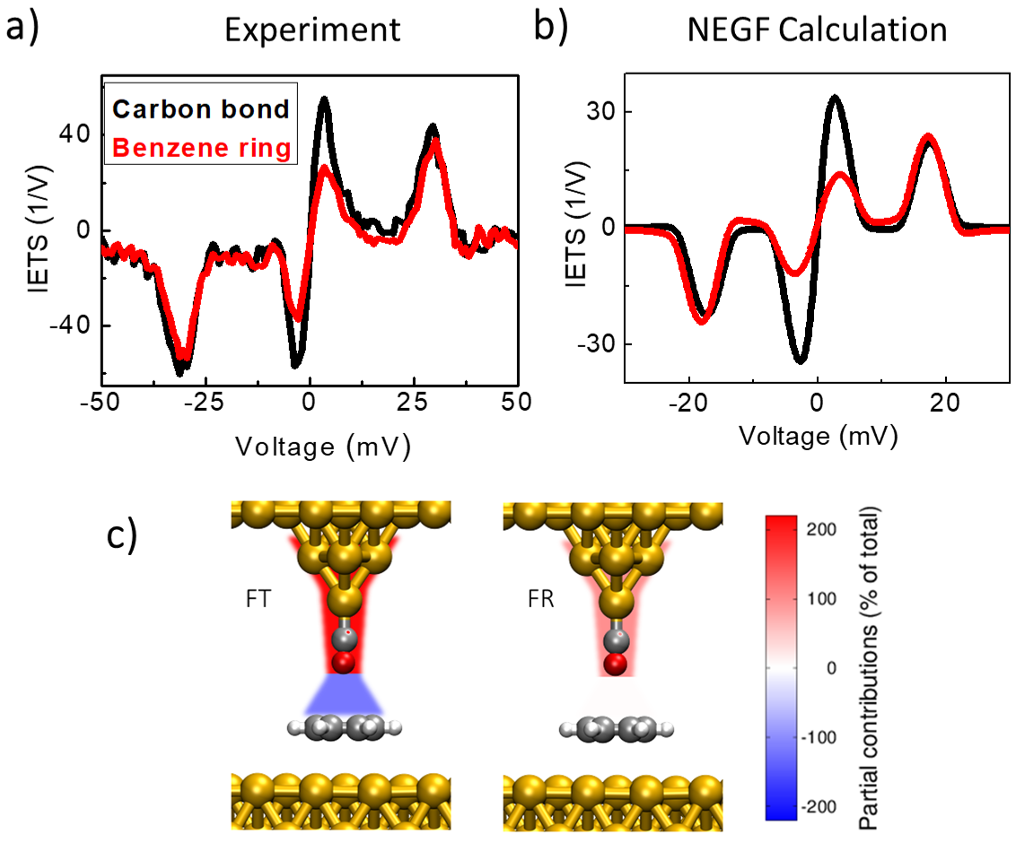}
\caption{\textbf{Detailed variation of IETS signal .} Experimental
(a) and calculated (b) IETS of the frustrated translational (FT)
and frustrated rotational (FR) modes above a carbon bond (black)
and a benzene ring (red). Both experimental and calculated plots
show the variation of the amplitude in the FT mode responsible of
the IETS contrast. c) Origin of the inelastic signal (as fraction
of the total inelastic amplitude) above the benzene ring. The
signal for FR modes originates mainly in the CO and tip. However,
for FT modes, the CO-benzene coupling also plays a significant
role.}
 \label{fig-03}
\end{figure}

For a further understanding of the origin of the contrast observed
in the IETS images, we analyze the IETS signal at two different
sites over one peripheral benzene ring of the FePc molecule. The
spectra were acquired at constant-height using the same parameters
for the lock-in as during the IETS image acquisition.
\reffig{fig-03}~a) shows two IETS curves, obtained on a carbon bond
(black) and at the center of the benzene ring (red) for
comparison. The IETS shows a strong variation of the amplitude of
the FT mode while the FR remains almost constant. This observation also documents that it is
not possible to clearly resolve the molecular structure using the
FR mode \cite{Chiang_Science2014}.

To get more insight into the imaging mechanism, we carry out first
principles calculations of the inelastic signal on a simplified
system consisting of a CO-functionalized tip placed above a
benzene molecule. As we show below, this system is already
sufficient to explain the experimental measurements. Details and
methodology are given in the Supplemental Information \cite{SOM}.

\reffig{fig-03}~b) shows the calculated IETS
spectrum for the CO-functionalized tip above a carbon bond and the
benzene ring. Consistently with the experiments, the intensity of
the calculated FT peak is reduced by a factor $\sim$2,  when going
from the carbon bond to the benzene ring, while the intensity of
the FR peaks is almost unchanged.

To understand why the inelastic signal of the FT modes changes strongly with tip position but not that of the FR modes, we analyze the different contributions that give rise to the inelastic peaks. The inelastic signal can be understood in terms of a Fermi golden rule involving the deformation potential and the left- and right-incident scattering states \cite{Paulsson_PropRules_2008}. The use of a local-orbital basis enables us to group the contributions of the various regions of the junction where the inelastic signal is generated \cite{Foti_LocalContrib_2016}. We consider a first set of contributions involving the CO molecule and the Au tip, and a second set consisting of the CO-benzene coupling (see \cite{SOM} for details). The effect of
the remaining terms is negligible. The modulus square of the sum of all contributions gives the calculated intensity of each inelastic peak.

These contributions for the most active pair of eigenchannels are
shown in \reffig{fig-03}~c) for the CO-functionalized tip above the
benzene ring. For clarity this is shown for only one FT and one FR
modes but results are similar for all FT and FR modes on both
configurations (details are given in the Supplemental Information \cite{SOM}).
The boxes in \reffig{fig-03}~c)  show the fraction of
the total inelastic signal arising from each set of contributions:
CO and tip, and the coupling of CO to the benzene substrate. The
color scale quantifies the magnitude of each set of contributions
relative to the total.

From the results shown above, we infer that there are clear differences in the origin of FT and FR modes. The FR mode is well described by the first set of contributions only. However, in order to properly capture the intensity of the FT mode, it is necessary to include these contributions as well as the off-diagonal terms between CO and benzene. Thus the inelastic signal of the FR mode is generated almost completely on the CO and tip alone, while that of the FT mode also involves the interaction with the benzene molecule.

The consequences of this localization of the inelastic signal are illustrated by constraining the dynamical region to just the C or O atom (see the Supplemental Information \cite{SOM}). When only the vibrations of the C atom are considered, the inelastic signal is unchanged with tip position. However, when only the O atom is allowed to vibrate, the calculated inelastic signal changes substantially from carbon bond to benzene ring positions. Thus the sensitivity of the inelastic peaks to the interaction with the molecular substrate rests on the composition of the vibrational modes of CO. In FR modes the larger displacement corresponds to the C atom, close to the tip. In FT modes, on the other hand, the displacement is larger on the O atom which, being closer to the benzene molecule, is more affected by the interaction with the molecular substrate. The higher sensitivity of FT modes to the position of CO above the molecule follows intuitively from this result.

To describe the variations of the IETS amplitude of the FT mode,
we extend the fast PP-IETS method to calculate IETS maps
\cite{Hapala_PRL2014}. We carry out an~approximation of the
perturbative inelastic transport theory
~\cite{Paulsson_PropRules_2008} and consider that the IETS signal
is proportional to two terms only: i) Variation of the tunnelling
hopping $T_{mn}$ between the electronic states of the tip ($m$)
and of the sample ($n$) with respect to displacement of PP along
vibration eigenmode $\mathbf{v}_{\lambda}$. ii) A~pre-factor
depending on the energy of the vibrational mode $\omega_\lambda$.
Intuitively the pre-factor represents the magnitude of the CO
displacement during vibration, which is prolonged as vibration
mode becomes softer due to concave potential over the bonds
\cite{Hapala_PRL2014}. The full derivation of the approach can be
found in the Supplemental Information \cite{SOM}.This
approximation is justified by the NEGF analysis, which revealed
dominant contribution of oxygen to the IETS signal of the FT mode
\cite{SOM}.
Consequently, the IETS signal $\gamma_{FT}$ of the FT
modes is defined as follows:
\begin{equation}
\gamma_{FT} \sim \sum_{\lambda,m,n} C \frac{1}{\omega_\lambda M_{PP}} \left| \frac{\partial T_{mn}}{\partial \mathbf{v}_{\lambda}} \right|^2,
\label{eq_IETS_main}
\end{equation}
where $C$ is a~constant and $M_{PP}$ the effective PP mass \cite{SOM}. The vibrational mode $\lambda$ goes over two FT modes. This PP-IETS method was implemented into the PP-code allowing simulation of the HR-STM and AFM images \cite{Krejci_PRB2017,Hapala_PRB2014,Hapala_PRL2014}. The computational cost of IETS images using this method is similar to standard STM simulation. The calculation of the high-resolution AFM/STM/IETS images relies on atomic and electronic structure of the adsorbed molecules on surface. Therefore, we performed total energy DFT simulations of FePc molecule on Au(111) surface with FHI-Aims code \cite{AIMS} using PBE functional \cite{PBE} and Tkatchenko-Scheffler vdW model \cite{Tkatchenko_2010} for geometry optimization. The electronic states used as input for PP-STM and PP-IETS codes were then calculated with hybrid B3LYP functional \cite{B3LYP_changed} to provide better description of the metal-organic system.

Electronic states of the CO-functionalized tip were approximated by $p_x$, $p_y$ and $s$ orbitals on the \PP{} to represent $\pi$ and $\sigma$  conductance channels \cite{Krejci_PRB2017}.
More details about parameters of the total energy DFT and PP-SPM simulations can be found in the Supplemental Information \cite{SOM}.

\begin{figure}
    \centering
    \includegraphics[width=8.5cm]{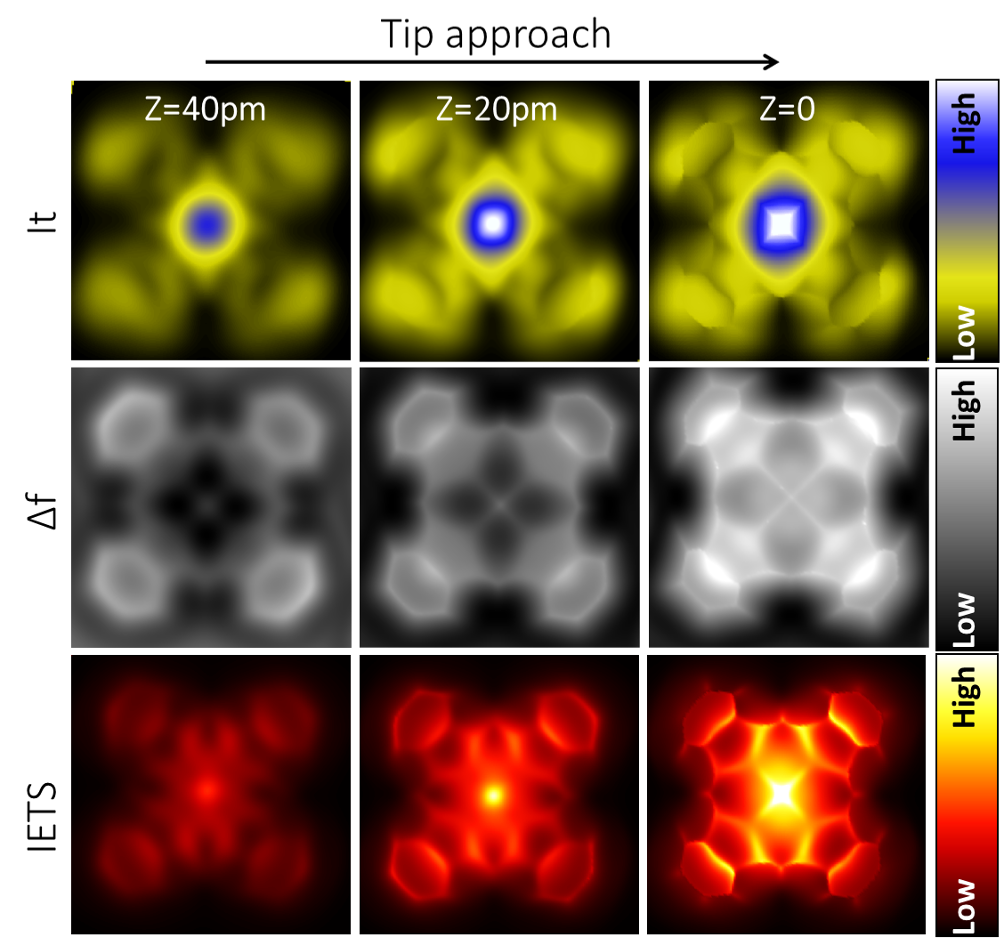}
    \caption{\textbf{Theoretical simulations of STM/AFM/IETS of FePc on Au(111) surface at three different tip-sample distances.} The tunnelling current (top), $\Delta$f (middle) and IETS (bottom) images were calculated using the PP-SPM model. }
    \label{fig-04}
\end{figure}

\reffig{fig-04} shows the calculated high-resolution AFM, STM and IETS images of FePc molecule on Au(111) surface for different tip-sample distances. The simulated images capture well most of the characteristic features observed experimentally in all channels. In STM mode, we are able to reproduce the dominant contrast observed in the center of the molecule, while a sharp contrast on external benzene rings is visible only in the close distance. On the other hand, the calculated STM images contain subtle submolecular features, which are not seen in the experiment. The AFM simulation reproduces very well both the characteristic sharpening of contrast and the contrast inversion at close distances. The only discrepancy consists of missing rectangular feature in the center of the FePc molecule. We attribute these discrepancies to a~peculiar charge distribution or structural relaxation between Fe and ligand, which is not described well within DFT approximation, or possibly to some chemical force between CO tip and Fe atom which cannot be captured using a Lennard-Jones force-field. The IETS channel resolves the molecular skeleton, with the characteristic bright spot in the center of the molecule in good agreement with the experimental evidence.

From PP-IETS simulations we can also analyze the influence of two contribution on the amplitude of the IETS signal (see Fig.~S13 in \cite{SOM}). We found that the pre-factor $1/{\omega}_{\lambda}$ (\refeq{eq_IETS_main}) is important only at close tip-sample distances due to fast decay of Pauli repulsion which leads to concave curvature of interaction potential over the bonds. However, the molecular structure is resolved, both in the theory and experimental, over range of $\approx$ 0.8~\AA (see Fig.~S1 in \cite{SOM}). The IETS contrast at far distance is mainly caused by spatial variation of inelastic tunnelling matrix element, as shown both by NEGF and PP-IETS simulations.

In conclusion, we demonstrated that the high-resolution IETS
imaging with CO-functionalized tip is feasible at 5 K with
a spatial resolution superior to STM and comparable to AFM. Thus
we believe that the IETS mode is promising for STM-only setups. The high-resolution contrast observed
in all channels simultaneously demonstrates unambiguously the
common imaging mechanism of the AFM/STM/IETS, related to lateral
bending of the CO-functionalized tip. We provided detailed
theoretical analysis of FT and FR vibrational modes showing their
different localization in the STM junction. This explains why
FT displays a large variation during scanning while FR is insensitive. This result opens the way to tune the surface
sensitivity of the inelastic signal through appropriate molecular
functionalization of the tip. We also showed that the submolecular
contrast emerges not only from the changes of the CO vibrational
frequency of the FT mode, but also due to the variation of the
amplitude of the IETS signal. Finally, we extended accordingly the
probe-particle AFM/STM/IETS model to include these two main
ingredients necessary to reproduce the high-resolution IETS
contrast.

\section{ acknowledgements }

This research was financially supported from the Czech Science Foundation (GA\v{C}R) under projects 15-19672S and 17-24210Y, the Purkyn\v{e} Fellowship program of the Academy of Sciences of the Czech Republic, and the European Union’s Horizon 2020 research and innovation programme under the Marie Sk\l{}odowska-Curie grant agreement No 709114. The authors also acknowledge support from Ministry of Education, Youth and Sport of the Czech Republic  (NanoEnviCz, LM2015073; and the project LO1305), FP7 FET-ICT “Planar Atomic and Molecular Scale devices” (PAMS) project (funded by the European Commission under contract No. 610446), Spanish Ministerio de Economia y Competitividad (MINECO) (Grant No. MAT2016-78293-C6-4-R) and the Basque Government (Dep. de Educacion and UPV/EHU, Grants No.IT-756-13 and  PI-2016-1-0027).
We thank CESNET LM2015042 and CERIT Scientific Cloud LM2015085, under the program ‘Projects of Large Research, Development, and Innovations Infrastructures’ for computational resources.


	%
\end{document}